\documentclass[10pt]{article}
\usepackage{amsmath, amsthm}

\newtheoremstyle{theorem}
{10pt} 
{10pt} 
{\sl} 
{\parindent} 
{\bf} 
{. } 
{ } 
{} 
\theoremstyle{theorem}

\newtheoremstyle{defi}
{10pt} 
{10pt} 
{\rm} 
{\parindent} 
{\bf} 
{. } 
{ } 
{} 
\theoremstyle{defi}

\begin{document}

\title{Family of non-equilibrium statistical operators and influence of the past on the present}
\author{V.V. Ryazanov\\
Institute for Nuclear Research, Kiev, pr.Nauki, 47 Ukraine\\
vryazan@kinr.kiev.ua}

\maketitle

\begin{abstract}
A family of non-equilibrium statistical operators (NSO) is
introduced which differ by the system lifetime distribution over
which the quasi-equilibrium (relevant) distribution is averaged.
This changes the form of the source in the Liouville equation, as
well as the expressions for the kinetic coefficients, average
fluxes, and kinetic equations obtained with use of NSO. It is
possible to choose a class of lifetime distributions for which
thermodynamic limiting transition and to tend to infinity of
average lifetime of system is reduced to the result received at
exponential distribution for lifetime, used by Zubarev. However
there is also other extensive class of realistic distributions of
lifetime of system for which and after to approach to infinity of
average lifetime of system non-equilibrium properties essentially
change. For some distributions the effect of "finite memory" when
only the limited interval of the past influence on behaviour of
system is observed. It is shown, how it is possible to spend
specification the description of effects of memory within the
limits of NSO method, more detailed account of influence on
evolution of system of quickly varying variables through the
specified and expanded form of density of function of distribution
of lifetime. The account of character of history of the system,
features of its conduct in the past, can have substantial
influence on non-equilibrium conduct of the system in a present
moment time.\\

{\bf AMS Subject Classification: 82C03; 82C70}\\

{\bf Key Words and Phrases: non-equilibrium statistical operator,
lifetime, account of character of history of the system}
\end{abstract}

\section{Introduction}

One of the most fruitful and successful ways of development of the
description of the non-equilibrium phenomena are served by a
method of the non-equilibrium statistical operator (NSO)
\cite{zub74,zub80}. In work \cite{ry01} new interpretation of a
method of the NSO is given, in which operation of taking of
invariant part \cite{zub74,zub80} or use auxiliary "weight
function" (in terminology \cite{ra95,ra99}) in NSO are treated as
averaging of quasi-equilibrium statistical operator on
distribution of past lifetime of system. This approach adjust with
the operations spent in the general theory of random processes, in
the renewal theory, and also with the lead Zubarev in work as
\cite{zub80} reception NSO by means of averaging on the initial
moment of time.

This treatment of NSO gives to the procedure looking before
formal, physical sense of the account of causality and allocation
of a real finite time interval in which there is a given physical
system. New interpretation leads to various directions of
development of NSO method which is compared, for example, with
Prigogine's \cite{Prig} approach, introduction of the operator of
internal time, irreversibility at microscopical level.

In Kirkwood's works \cite{kirk} it was noticed, that the system
state in time present situation depends on all previous evolution
of the non-equilibrium processes developing it. For example, in
real crystals it is held in remembrance their formation in various
sorts "defects" (dispositions etc.), reflected in structure of the
crystals. Changing conditions of formation of crystals, we can
change their properties and create new materials. In works
\cite{ra95,ra99} it is specified, that it is possible to use many
"weight functions". Any form of density of lifetime distribution
gives a chance to write down a source of general view in dynamic
Liouville equation which thus becomes, specified Boltzmann and
Prigogine \cite{ra95,ra99,Prig}, and contains dissipative items.

If in Zubarev's works \cite{zub74,zub80} the linear form of a
source corresponding limiting exponential distribution for
lifetime is used other expressions for density of lifetime
distribution give fuller and exact analogues of "integrals of
collisions". The obvious account of violation of time symmetry
(through finiteness of lifetime, the beginning, the end and
irreversibility of a life) is entered. Besides communication with
the theory of queues, reliability theory, the management theory,
the information theory etc., in offered work the physical
consequences connected with fundamental physical problems are
reflected.

The formalism follows from the physical matter, for example, from
finiteness of lifetime of real physical systems (it is possible to
result many examples of problems in which it is necessary to
consider systems of the finite sizes with finite lifetime).
Generally the description of non-equilibrium systems represents
the self-coordinated problem: definition of lifetime through
interaction of system with environment \cite{zub74}, dynamics of
the operators characterizing non-equilibrium processes, and
substitution of found average lifetime in NSO, definition of
non-equilibrium physical characteristics, depending on system
lifetime.

In work \cite{rau} irreversible transfer equations are received in
assumptions of coarsening of the distributions, a certain choice
of macroscopical variables and the analysis of division of time
scales of the description (last circumstance was marked in
\cite{bog}). Importance and necessity of the analysis of the time
scales playing a fundamental role in the description of
macroscopical dynamics of system is underlined. Evolution of slow
degrees of freedom is described by Markovian equations. Thus the
time scale on which observable variables evolve, should be much
more time of memory on which the residual effects brought by
irrelevant degrees of freedom are considered.

Otherwise effects of memory play an essential role. Memory time is
estimated in work \cite{rau} for Boltzmann equation. In the
present work the consideration subject is made by situations when
it is necessary to consider effects of memory. Examples of such
situations are given in \cite{rau}. We will notice, that the
projective methods used in \cite{rau}, do not consider
distribution on lifetime of system (that is noted in \cite{zub80})
on which method NSO is based.

In work \cite{ry07} it is shown, in what consequences for
non-equilibrium properties of system results change of lifetime
distribution of system for systems of the limited volume with
finite lifetime. In the present work are considered also
infinitely greater systems with infinite average lifetime.

\section{New interpretation of NSO}
\label{sect:1}

In \cite{ry01} the Nonequilibrium Statistical Operator introduced
by Zubarev \cite{zub74,zub80} rewritten as

$$
ln\varrho(t)= \int_{0}^{\infty}p_{q}(u)ln\varrho_{q}(t-u,
-u)du,\quad
ln\varrho_{q}(t, 0)=-\Phi(t)-\sum_{n}F_{n}(t)P_{n};\\
$$
$$
ln\varrho_{q}(t, t_{1})=e^{\{-t_{1}H/i\hbar\}}ln\varrho_{q}(t,
0)e^{\{t_{1}H/i\hbar\}}; \quad \Phi(t)=\ln Sp
\exp\{\sum_{n}F_{n}(t)P_{n}\},
$$
where $H$ is hamiltonian, $ln\varrho(t)$ is the logarithm of the
NSO in Zubarev's form, $ln\varrho_{q}(t, 0)$ is the logarithm of
the quasi-equilibrium (or relevant); the first time argument
indicates the time dependence of the values of the thermodynamic
parameters $F_{m}$; the second time argument $t_{2}$ in
$\varrho_{q}(t_{1}, t_{2})$ denotes the time dependence through
the Heizenberg representation for dynamical variables $P_{m}$ from
which $\varrho_{q}(t, 0)$ can depend
\cite{zub74,zub80,ry01,ra95,ra99}. In \cite{ry01} the auxiliary
weight function $p_{q}(u)=\varepsilon exp\{-\varepsilon u\}$ was
interpreted as the probability distribution of lifetime density of
a system. $\Gamma$ is random variables of lifetime from the moment
$t_{0}$ of its birth till the current moment $t$;
$\varepsilon^{-1}=\langle t-t_{0}\rangle$; $\langle
t-t_{0}\rangle=\langle\Gamma\rangle$, where
$\langle\Gamma\rangle=\int u p_{q}(u)du$ is average lifetime of
the system. This time period can be called the time period of
getting information about system from its past. Instead of the
exponential distribution $p_{q}(u)$ in (1) any other sample
distribution could be taken. This fact was marked in \cite{ry01}
and \cite{ra95,ra99} (where the distribution density $p_{q}(u)$ is
called auxiliary weight function $w(t,t`)$). From the complete
group of solutions of Liouville equation (symmetric in time) the
subset of retarded "unilateral" in time solutions is selected by
means of introducing a source in the Liouville equation

$$
\frac{\partial\varrho(t)}{\partial
t}+iL\varrho(t)=-\varepsilon(\varrho(t)-\varrho_{q}(t, 0))=J,
$$
which tends to zero (value $\varepsilon\rightarrow 0$) after
thermodynamic limiting transition. Here $L$ is Liouville operator;
$iL=-\{H,\varrho\}=\Sigma_{k}[\frac{\partial H}{\partial
p_{k}}\frac{\partial\varrho}{\partial q_{k}}-\frac{\partial
H}{\partial q_{k}}\frac{\partial\varrho}{\partial p_{k}}]$; $H$ is
Hamilton function, $p_{k}$ and $q_{k}$ are pulses and coordinates
of particles; $\{...\}$ is Poisson bracket. In \cite{mor} it was
noted that the role of the form of the source term in the
Liouville equation in NSO method has never been investigated. In
\cite{fe} it is stated that the exponential distribution is the
only one which possesses the Markovian property of the absence of
contagion, that is whatever is the actual age of a system, the
remaining time does not depend on the past and has the same
distribution as the lifetime itself. It is known
\cite{zub74,zub80,ry01,ra95,ra99} that the Liouville equation for
NSO contains the source $J=J_{zub}=-\varepsilon [ln\varrho
(t)-ln\varrho_{q}(t,0)]$ which becomes vanishingly small after
taking the thermodynamic limit and setting $\varepsilon\rightarrow
0$, which in the spirit of the paper \cite{zub74} corresponds to
the infinitely large lifetime value of an infinitely large system.
For a system with finite size this source is not equal to zero. In
\cite{ra99} this term enters the modified Liouville operator and
coincides with the form of Liouville equation suggested by
Prigogine \cite{Prig} (the Boltzmann-Prigogine symmetry), when the
irreversibility is entered in the theory on the microscopic level.
We note that the form of NSO by Zubarev cast in \cite{ry01}
corresponds to the main idea of \cite{Prig} in which one sets to
the distribution function $\varrho$ ($\varrho_{q}$ in Zubarev's
approach) which evolves according to the classical mechanics laws,
the coarse distribution function $\widetilde{\varrho}$
($\varrho(t)$ in the case of Zubarev's NSO) whose evolution is
described probabilistically since one perform an averaging with
the probability density $p_{q}(u)$. The same approach (but instead
of the time averaging the spatial averaging was taken) was
performed in \cite{klim}.

Besides the Zubarev's form of NSO \cite{zub74,zub80}, NSO
Green-Mori form \cite{gr,mo} is known, where one assumes the
auxiliary weight function \cite{ra95} to be equal
$W(t,t')=1-(t-t')/\tau; w(t,t')=dW(t,t')/dt'=1/\tau;
\tau=t-t_{0}$. After averaging one sets $\tau\rightarrow\infty$.
This situation at $p_{q}(u=t-t_{0})=w(t,t'=t_{0})$ coincides with
the uniform lifetime distribution. The source in the Liouville
equation takes the form $J=ln\varrho_{q}/\tau$. In \cite{zub74}
this form of NSO is compared to the Zubarev's form.

One could name many (no less than 1000) examples of explicit
defining of the function $p_{q}(u)$. Every definition implies some
specific form of the source term $J$ in the Liouville equation,
some specific form of the modified Liouville operator and NSO.
Thus the family of NSO is defined. If distribution $p_{q}(u)$
contains $n$ parameters, it is possible to write down n equations
for their expression through the parameters of the system. From
other side, they are expressed through the moments of lifetime.
There is the problem of optimum choice of function $p_{q}(u)$ and
NSO.

\section{Modifications to the nonequilibrium description}
\label{sect:2}

Let's consider now, what consequences follow from such
interpretation of NSO.

\subsection{Families of NSO}
\label{subsect:A}

Setting various distributions for past lifetime of the system, we
receive a way of recording of families of NSO. Class of NSO from
this family will be connected with a class of distributions for
lifetime (taken, for example, from the stochastic theory of
storage processes, the theory of queues etc.) and with relaxation
properties of that class of physical systems which is
investigated. The general expression for NSO with any distribution

\begin{equation}
ln\varrho(t)= \int_{0}^{\infty}p_{q}(u)ln\varrho_{q}(t-u,
-u)du=
\label{NSO}
\end{equation}

$$
=ln\varrho_{q}(t, 0)-\int_{0}^{\infty}(\int p_{q}(u)du)\frac{d
\ln\varrho(t-u, -u)}{du}du,
$$
where integration by parts in time is carried out at  $\int p_{q}
(y)dy_{|y=0}=-1; \\ \int p_{q}(y)dy_{|y\rightarrow\infty}=0$; at
$p_{q}(y)=\varepsilon\exp\{-\varepsilon y\};
\varepsilon=1/\langle\Gamma\rangle$, the expression (1) passes in
NSO from \cite{zub74,zub80}. In \cite{fe} it is shown, how from
random process ${X (t)}$, corresponding to evolution of
quasi-equilibrium system, it is possible to construct set of new
processes, introducing the randomized operational time. It is
supposed, that to each value $t>0$ there corresponds a random
value $\Gamma(t)$ with the distribution $p^{t}_{q}(y)$. The new
stochastic kernel of distribution of a random variable
$X(\Gamma(t))$ is defined by equality of a kind (1). Random
variables $X(\Gamma(t))$ form new random process which, generally
speaking, need not to be of Markovian type any more. Each moment
of time $t$ of "frozen" quasi-equilibrium system is considered as
a random variable $\Gamma(t)$ the termination of lifetime with
distribution $p^{t}_{q}(y)$. Any moment of lifetime can be with
certain probability the last. That the interval $t-t_{0}=y$ was
enough large (that became insignificant details of an initial
condition as dependence on the initial moment $t_{0}$ is
nonphysical \cite{zub74,zub80}), it is possible to introduce the
minimal lifetime $\Gamma_{min}=\Gamma_{1}$ and to integrate in
(\ref{NSO}) on an interval $(\Gamma_{1},\infty)$. It results to
the change of the normalization density of distribution
$p_{q}(y)$. For example, the function $p_{q}(y)=\varepsilon\exp
\{-\varepsilon y\}$ will be replaced by $p_{q}(y)=\varepsilon\exp
\{\varepsilon\Gamma_{1}-\varepsilon y\}, y\geq\Gamma_{1};
p_{q}(y)=0, y < \Gamma_{1}$. The under limit of integration in
(\ref{NSO}) by $\Gamma_{1}\rightarrow 0$ is equal $0$. It is
possible to choose $p_{q}(y)=Cf(y), y<t_{1};  p_{q}(y)
=\varepsilon\exp \{-\varepsilon y\}, y\geq t_{1};
C=(1-exp\{-\varepsilon t_{1}\})/(\int_{0}^{t_{1}} f(y)dy)$. The
function $f(y)$ can be taken from models of the theory of queues,
the stochastic theory of storage and other sources estimating the
lifetime distribution for small times (for example
\cite{co,cox,str61,tur,kor}). The value $t_{1}$ can be found from
results of work \cite{str61}. It is possible to specify many
concrete expressions for lifetime distribution of system, each of
which possesses own advantages. To each of these expressions there
corresponds own form of a source in Liouville equation for the
nonequilibrium statistical operator. In general case any functions
$p_{q}(u)$ the source is:

\begin{equation}
J=p_{q}(0)\ln\varrho_{q}(t, 0)+\int_{0}^{\infty}\frac{\partial
p_{q}(y)}{\partial y}(ln\varrho_{q}(t-y, -y))dy \label{sour}
\end{equation}
(when values $p_{q}(0)$ disperse, it is necessary to choose the
under limit of integration equal not to zero, and $\Gamma_{min}$).
Such approach corresponds to the form of dynamic Liouville
equation in the form of Boltzmann-Bogoliubov-Prigogine
\cite{ra95,ra99,Prig}, containing dissipative items.

Thus the operations of taking of invariant part \cite{zub74},
averaging on initial conditions \cite{zub80}, temporary
coarse-graining \cite{kirk}, choose of the direction of time
\cite{ra95,ra99}, are replaced by averaging on lifetime
distribution.

The physical sense of averaging on introduced lifetime
distribution of quasi-equilibrium system as it was already marked,
consists in the obvious account of infringement of time symmetry
and loss (reduction accessible) the information connected with
this infringement, that is shown in occurrence  the value of
average of entropy production $\langle\Delta S(t)\rangle$ not
equal to zero, obviously reflecting fluctuation-dissipative
processes at the real irreversible phenomena in non-equilibrium
systems. The correlations received in the present section
generalize results of statistical non-equilibrium thermodynamics
\cite{zub74,zub80} and information statistical thermodynamics
\cite{ra95,ra99} as instead of weight function of a form
$\varepsilon\exp\{\varepsilon t'\}$ contain density of probability
of lifetime of quasi-equilibrium system which as it was already
marked, can not coincide with exponential distribution (in the
latter case it coincides with weight function from
\cite{zub74,zub80}). For example, for system with $n$ classes of
ergodic states limiting exponential distribution is replaced with
the general Erlang. In research of lifetimes of complex systems it
is possible to involve many results of the theory of reliability,
the theory of queues, the stochastic theory of storage processes,
theory of Markov renewal, the theory of semi-Markov processes etc.

It is essentially that $\varepsilon\neq 0$. The thermodynamic
limiting transition is not performed, and actually important for
many physical phenomena dependence on the size of system are
considered. We assume $\varepsilon$ and $\langle\Gamma\rangle$ to
be finite values. Thus the Liouville equation for $\varrho(t)$
contains a finite source. The assumption about finiteness of
lifetime breaks temporary symmetry. And such approach
(introduction $p_{q}(y)$, averaging on it) can be considered as
completing the description of works \cite{zub74,zub80}.

In work \cite{str61} lifetimes of system are considered as the
achievement moments by the random process characterizing system,
certain border, for example, zero. In \cite{str61} are received
approached exponential expressions for density of probability of
lifetime, accuracy of these expressions is estimated. In works
\cite{str95,str96} lifetimes of molecules are investigated, the
affinity of real distribution for lifetime and approached
exponential model is shown. It is possible to specify and other
works (for example \cite{gasp93,gasp95,dorf}) where physical
appendices of concept of lifetime widely applied in such
mathematical disciplines, as reliability theory, the theory of
queues and so forth (under names non-failure operation time, the
employment period, etc.) are considered. In the present section
lifetime joins in a circle of the general physical values, acting
in an estimation or management role (in terminology of the theory
of the information \cite{str66}) for the quasi-equilibrium
statistical operator that allows to receive the additional
information on system. In \cite{str66} it is noticed, that three
disciplines grow together: statistical thermodynamics, Shennon's
theory of the information and the theory of optimum statistical
decisions. Accordingly, all correlations written down in the
present work can be interpreted in terms of the theory of the
information or the theory of optimum statistical decisions
\cite{tsen}.

Let's notice, that in a case when value
$d\ln\varrho_{q}(t-y,-y)/dy$ (the operator of entropy production
$\sigma$ \cite{zub74}) in the second item of the right part
(\ref{NSO}) does not depend from $y$ and is taken out from under
integral on $y$, this second item becomes
$\sigma\langle\Gamma\rangle$, and expression (\ref{NSO}) does not
depend on form of function $p_{q}(y)$. There is it, for example,
at $\varrho_{q}(t)\sim \exp\{-\sigma t\}, \sigma=const$. In work
\cite{dew}) such distribution is received from a principle of a
maximum of entropy at the set of average values of fluxes.

\subsection{Physical sense of distributions for past
lifetime of system} \label{subsect:B}

As is known (for example, \cite{co}, \cite{str61}, \cite{tur}),
exponential distribution for lifetime

\begin{equation}
p_{q}(y) =\varepsilon\exp\{-\varepsilon  y\}, \label{expDi}
\end{equation}
used in Zubarev's works \cite{zub74,zub80}, is limiting
distribution for lifetime, fair for large times. It is marked in
works \cite{zub74,zub80} where necessity of use of large times
connected with damping of nonphysical initial correlations. Thus,
in works \cite{zub74,zub80} the thermodynamic result limiting and
universal is received, fair for all systems. It is true in a
thermodynamic limit, for infinitely large systems. However real
systems have the finite sizes. Therefore essential there is use of
other, more exact distributions for lifetime. In this case the
unambiguity of the description peculiar to a thermodynamic limit
\cite{mart} is lost.

For NSO with Zubarev's function (\ref{expDi}) the value enter in
second item

\begin{equation}
-\int p_{q}(y)dy =\exp\{-\varepsilon  y\}=1-\varepsilon
y+(\varepsilon
y)^{2}/2-...=1-y/\langle\Gamma\rangle+y^{2}/2\langle\Gamma\rangle^{2}-...
\label{sum}
\end{equation}
Obviously that to tend to infinity of average lifetime,
$\langle\Gamma\rangle\rightarrow\infty$, correlation (\ref{sum})
tends to unity.

Besides exponential density of probability (\ref{expDi}), as
density of lifetime distribution Erlang distributions (special or
the general), gamma distributions etc. (see \cite{co,cox}), and
also the modifications considering subsequent composed asymptotic
of the decomposition \cite{tur} can be used. General Erlang
distributions for $n$ classes of ergotic states are fair for cases
of phase transitions or bifurcations. For $n=2$ general Erlang
distribution looks like
$p_{q}(y)=\theta\rho_{1}\exp\{-\rho_{1}y\}+(1-\theta)\rho_{2}\exp\{-\rho_{2}y\},
\theta<1$. Gamma distributions describe the systems which
evolution has some stages (number of these stages coincides with
gamma distribution order). Considering real-life stages in
non-equilibrium systems (chaotic, kinetic, hydrodynamic, diffusive
and so forth), it is easy to agree, first, with necessity of use
of gamma distributions of a kind

\begin{equation}
p_{q}(y) =\varepsilon(\varepsilon y)^{k-1}\exp\{-\varepsilon
y\}/\Gamma(k) \label{gam}
\end{equation}
($\Gamma(k)$ is gamma function, at $k=1$ we receive distribution
(\ref{expDi})), and, secondly, - with their importance in the
description of non-equilibrium properties.

More detailed description $p_{q}(u)$ in comparison with limiting
exponential (\ref{expDi}) allows to describe more in detail real
stages of evolution of system (and also systems with small
lifetimes). Each from lifetime distributions has certain physical
sense. In the theory of queues, for example \cite{prab}, to
various disciplines of service there correspond various
expressions for density of lifetime distribution. In the
stochastic theory of storage \cite{prab}, to these expressions
there correspond various models of an exit and an input in system.

The value $\varepsilon$ without taking into account of dissipative
effects can be defined, for example, from results of work
\cite{str61}. The value $\varepsilon$ is defined also in work
\cite{ry01} through average values of operators of entropy and
entropy production, flows of entropy and their combination.

How was already marked, it is possible to specify very much, no
less than $1000$, expressions for the distributions of past
lifetime of the system. Certain physical sense is given to each of
these distributions. To some class functions of distributions,
apparently, some class of the physical systems corresponds, the
laws of relaxation in which answer this class of functions of
distributions for lifetime.

\subsection{Influence of the past on
non-equilibrium properties} \label{subsect:C}

A). Expressions for average fluxes.

In \cite{fe} by consideration of the paradox connected with a
waiting time, the following result is received: let $X_{1}=S_{1};
X_{2}=S_{2}-S_{1}; ...$ are mutually independent also it is
equally exponential the distributed values with average
$1/\varepsilon$. Let $t>0$ is settled, but it is any. Element
$X_{k}$, satisfying to condition $S_{k-1}<t  \leq S_{k}$, has
density $\nu_{t}(x)=\varepsilon^{2}x\exp\{-\varepsilon x\}, 0<x
\leq t; \nu_{t}(x)=\varepsilon(1+\varepsilon x)\exp\{-\varepsilon
x\}, x>t$. In Zubarev's NSO \cite{zub74,zub80} the value of
lifetime to a present moment t, belonging lifetime $X_{k}$,
influence of the past on the present is considered. Therefore the
value $p_{q}(u)$ should be chosen not in the form of exponential
distribution (\ref{expDi}), and in a form

\begin{equation}
p_{q}(y) =\varepsilon^{2}y\exp\{-\varepsilon  y\}, \label{exp2}
\end{equation}
that in the form of gamma distribution (\ref{gam}) at $k=2$. In
this case distribution (\ref{exp2}) coincides with special Erlang
distribution of order $2$ \cite{co}, when refusal (in this case -
the moment $t$) comes in the end of the second stage \cite{co},
the system past consists of two independent stages. Function of
distribution is equal $P_{q}(x)=1-\exp\{-\varepsilon
x\}-\varepsilon x\exp\{-\varepsilon x\}, p_{q}(u)=dP_{q}(u)/du$,
unlike exponential distribution, when
$P_{q}(x)=1-\exp\{-\varepsilon x\}$. The behaviour of these two
densities of distribution of a form (\ref{expDi}) and (\ref{exp2})
essentially differs in a zero vicinity. In case of (\ref{exp2}) at
system low probability to be lost at small values $y$, unlike
exponential distribution (\ref{expDi}) where this probability is
maximal. Any system if has arisen, exists any minimal time, and it
is reflected in distribution (\ref{exp2}).

In work \cite{ry07} it is shown, in what differences from
Zubarev's distribution (\ref{NSO}) with exponential distribution
of lifetime (\ref{expDi}) results gamma distribution (\ref{gam}),
(\ref{exp2}) use. Additional items in NSO, in integral of
collisions of the generalized kinetic equation, in expressions for
average fluxes and self-diffusions coefficient are considered. The
same in [10] is done and for special Erlang distribution
$k=2,3,4..., n, P_{q}(x)=1-\exp\{-\varepsilon x\}[1+\varepsilon
x/1! + ... +(\varepsilon x)^{k-1}/(k-1)!];
\varepsilon=k/\langle\Gamma\rangle$. For distributions
(\ref{gam}), (\ref{exp2}) is correct correlation (4), value $-\int
p_{q}(u)du \rightarrow 1$ by
$\langle\Gamma\rangle\rightarrow\infty$.

Thus the multi-stage model of the past of system is introduced.
Non-equilibrium processes usually proceed in some stages, each of
which is characterized by the time scale. In distribution
(\ref{exp2}) the account of two stages, possibly, their minimal
possible number is made. Other distributions can describe any
other features of the past. Corresponding additives will be
included into expressions for fluxes, integral of collisions,
kinetic coefficients. Besides special Erlang distributions with
whole and specified values $k=n$, that does not deduce us from set
of one-parametrical distributions, the general already
two-parametrical gamma distribution where the parameter $k$ can
accept any values can be used. In this case $\langle\Gamma\rangle\
=k/\varepsilon$. The situation (formally), when $k<1$ is possible.
Then sources will tends to infinity, as
$(t-t_{0})^{k-1}_{|t\rightarrow t_{0}}\rightarrow\infty$ at $k<1$.
This divergence can be eliminated, having limited from below the
value $t-t_{0}$ of minimal lifetime value $\Gamma_{min}$, having
replaced the under zero limit of integration on $\Gamma_{min}$.
Then to expression for a source (\ref{sour}) it is added item
$[(\varepsilon
\Gamma_{min})^{k-1}/\Gamma(k)]\varepsilon\exp\{-\varepsilon
\Gamma_{min}\}\ln\varrho_{q}(t - \Gamma_{min}, - \Gamma_{min})$.

B). Entropy production.

Expressions for average entropy production received, for example,
in \cite{ra95}, also depend from $w$ (or $p_{q}(u)$ - in
designations \cite{ry01} and this work), i.e. on the chosen form
of density of probability of distribution of the past of system.
So, for average entropy production
$\overline{\sigma}=d\overline{S}/dt$  in work [4] expression
$\overline{\sigma}(t)=\Sigma_{k=1}^{\infty}\int_{t_{0}}^{t}dt'W(t,t')(\sigma(z|t,0);
\sigma(z|t',t'-t)|t)^{k}$ is received, where $z$ are points in
phase space , $\sigma(z|t',t'-t)=-d\ln\varrho_{q}(z|t',t'-t)/dt',
(\sigma(z|t,0); \sigma(z|t',t'-t)|t)^{k}=(k!)^{-1}\int
dz\sigma(z|t,0)\sigma(z|t',t'-t)\int_{t_{0}}^{t}dt_{1}W(t,t_{1})\sigma
(z|t_{1},t_{1}-t)...\int_{t_{0}}^{t} dt_{k-1}W(t,t_{k-1})\\
\sigma(z| t_{k-1},t_{k-1}-t)\varrho_{q}(z|t,0),
w(t,t')=dW(t,t')/dt'$ is "auxiliary weight function" (in
terminology [4]); $w (t, t')$ it is designated above as $p_{q}(u);
w(t, t')=p_{q}(u=t-t')$. For the limiting exponential distribution
(\ref{expDi}) used in Zubarev's NSO, $W(t,t')=\exp\{\varepsilon
(t'-t)\}$.

\section{Systems with infinite lifetime}
 \label{sect:3}

Above, as well as in work \cite{ry07}, additives to NSO in the
Zubarev's form are received for systems of the finite size, with
finite lifetime. We will show, as for systems with infinite
lifetime, for example, for systems of infinite volume, after
thermodynamic limiting transition, the same effects, which essence
in influence of the past of system, its histories, on its present
non-equilibrium state are fair.

In work \cite{ry07} it is shown, as changes in function $p_{q}(u)$
influences on non-equilibrium descriptions of the system. But for
those distributions $p_{q}(u)$, which are considered in
\cite{ry07} (gamma-distributions, (\ref{gam}), (\ref{exp2})) the
changes show up only for the systems of finite size with finite
lifetimes. Additions to unit in equation (\ref{sum}) becomes
vanishingly small to tend to infinity of sizes of the system and
its average lifetime, as in the model distribution (\ref{expDi})
used in Zubarev's NSO (\ref{NSO}). For the systems of finite size
and the exponential distribution results to nonzero additions in
expression (\ref{NSO}). Thus, these additions to NSO and proper
additions to kinetic equations, kinetic coefficients and other
non-equilibrium descriptions of the system, are an effect
finiteness of sizes and lifetime of the system, not choice of
distribution of lifetime of the system. We will find out, whether
there are distributions of lifetime of system for which and for
the infinitely large systems with infinitely large lifetime an
additional contribution to NSO differs from Zubarev'a NSO.

4a). Let's consider in quality $p_{q}(u)$ distribution of a form

$$
p_{q}(u)=\frac{ku^{k-1}\rho^{k}}{[1+(u\rho)^{k}]^{2}},
$$
received in work \cite{cox}, where $k=1/\tau, \nu=-\log \rho,
\tau$ and $\nu$ are parameters of scale and shift of logistical
distribution
$f(x)=\tau^{-1}\exp[(x-\nu)/\tau]/\{1+\exp[(x-\nu)/\tau]\}^{2}$.
In the correlatioon (\ref{NSO}) the value $\int p_{q}(u)du
=-1/[1+(u\rho)^{k}]$ appears. Average value of lifetime is equal

\begin{equation}
\langle\Gamma\rangle=\int_{0}^{\infty}u
p_{q}(u)du=\rho^{-1}B(1/(k+1),1-1/k), \label{av}
\end{equation}
$$
\langle\Gamma\rangle^{2}=\int_{0}^{\infty}u^{2}
p_{q}(u)du=\rho^{-2}B(2/(k+1),1-2/k),
$$
where $B( , )$ is beta function [32].

The value $\langle\Gamma\rangle$ in (\ref{av}) to tend to infinity
at $a)  \rho=0, b)  k=1$.  Ratio of the second moment toward the
square of the first is equal

\begin{equation}
\frac{\langle\Gamma^{2}\rangle}{\langle\Gamma\rangle^{2}}=
\frac{B(2/(k+1),1-2/k)}{B^{2}(1/(k+1),1-1/k)}. \label{ra}
\end{equation}

Expression (\ref{ra}) becomes vanishingly small with
$k\rightarrow1$. Correlation (\ref{NSO}) for this distribution
takes the form

$$
\ln\varrho(t)= \int_{0}^{\infty}p_{q}(y)\ln\varrho_{q}(t-y,
-y)dy=ln\varrho_{q}(t, 0)+ \\
$$
$$
+\int_{0}^{\infty}(\frac{1}{[1+(u\rho)^{k}]})(\frac{d\ln\varrho_{q}(t-u,
-u)}{du})du.
$$
At $k\rightarrow1$ and finite values $\rho$ we have a difference
from the zero of additions to unit in expansion

$$
\ln\varrho(t)=ln\varrho_{q}(t, 0)+
\int_{0}^{\infty}(1-(u\rho)^{k}+(u\rho)^{2k}-(u\rho)^{3k}+...)(d\ln\varrho_{q}(t-u,
-u)/du)du.
$$

But value $k$ it is possible to define from correlation
(\ref{ra}), and, if (\ref{ra}) is finite value not equal to the
zero then $k\neq1$. There is $\rho\rightarrow0$, when additions
becomes vanishingly small, as in the case of (\ref{sum}) of
Zubarev's NSO.

4b). Pareto distribution \cite{cox}

\begin{equation}
p_{q}(u)=\frac{ka^{k}}{[u+a]^{k+1}}=\frac{k(k/\rho_{0})^{k}}{[u+(k/\rho_{0})]^{k+1}},\quad
\rho_{0}=k/a. \label{Pa}
\end{equation}
This distribution is received in \cite{cox}, as complex
exponential distribution. It is supposed, that intensity $\rho$ of
exponential distribution $f(u)=\rho\exp\{-\rho u\}$ represents
random variable $P$ with distribution $f_{P}(\rho)$. Then

$$
  f_{T}(u)=p_{q}(u)=\int_{0}^{\infty}\rho\exp\{-\rho
  u\}f_{P}(\rho)d\rho.
$$
If to be set for function $f_{P}(\rho)$ by gamma distribution with
density

\begin{equation}
f_{P}(\rho)=a^{k}\rho^{k-1}\exp\{-a\rho\}/\Gamma(k) \label{ga}
\end{equation}
that we receive distribution (\ref{Pa}), Pareto distribution. From
(\ref{Pa}) it is obtained:

$$
\int
p_{q}(u)du=-\frac{a^{k}}{(u+a)^{k}}=-[1-\frac{uk}{a}+\frac{u^{2}k(k+1)}{2a^{2}}-\frac{u^{3}k(k+1)(k+2)}{6a^{3}}+...];\\
$$
$$
\langle\Gamma\rangle=a/(k-1), \quad k\geq1.
$$

We will consider two cases:

a). The parameter $k$ is fixed, $a=\langle\Gamma\rangle(k-1)$.
Then, as in case of exponential (\ref{expDi}) or gamma
distributions (\ref{gam}) for $p_{q} (u)$ additives to NSO are
proportional to $1/\langle\Gamma\rangle$, they becomes vanishingly
small with $\langle\Gamma\rangle\rightarrow\infty$ as in
(\ref{sum}).

b). The parameter a is fixed. Then $k=1+a/\langle\Gamma\rangle$;
$-\int
p_{q}(u)du=a^{k}/(u+a)^{k}=1-(u/a)(1+a/\langle\Gamma\rangle)
+(u^{2}/2a^{2})(1+a/\langle\Gamma\rangle)(2+a/\langle\Gamma\rangle)
-(u^{3}/6a^{3})(1+a/\langle\Gamma\rangle)(2+a/\langle\Gamma\rangle)
(3+a/\langle\Gamma\rangle)+ ... \rightarrow1/(1+u/a)$,
$\langle\Gamma\rangle\rightarrow\infty, k\rightarrow1$. We will
mark that the second moment of Pareto distribution does not exist
at $k \leq 2$ \cite{kor}. And at
$\langle\Gamma\rangle\rightarrow\infty, k\rightarrow1$.

Pareto distribution (\ref{Pa}) corresponds to Tsallis distribution
\cite{tsal}

$$
p_{q}(u)=\frac{1}{Z[1+\beta(q-1)u]^{1/(q-1)}}; \quad
k+1=\frac{1}{(q-1)}; \quad q=\frac{(k+2)}{(k+1)}; \quad
a=\frac{(k+1)}{\beta}.
$$
In  Tsallis method averaging is conducted on
distribution

$$
p^{q}(u)=\frac{(k/a)^{(k+2)/(k+1)}}{[1+u/a]^{k+2}}.
$$
Then

$$
\langle\Gamma\rangle=\frac{\int_{0}^{\infty}up^{q}(u)du}{\int_{0}^{\infty}p^{q}(u)du}=\frac{a}{k};
\quad  \frac{D}{\langle\Gamma^{2}\rangle}=\frac{(k+1)}{(k-1)};
\quad  D=\langle\Gamma^{2}\rangle-\langle\Gamma\rangle^{2};
$$
$$
k=\frac{(s+1)}{(s-1)}; \quad s=\frac{D}{\langle\Gamma^{2}\rangle};
\quad  -\int
p_{q}(u)du=\frac{a^{k}}{(u+a)^{k}}=\\
$$
$$
=1-\frac{uk}{a}+\frac{u^{2}k(k+1)}{2a^{2}}-\frac{u^{3}k(k+1)(k+2)}{6a^{3}}+...=\\
$$
$$
=1-\frac{u}{\langle\Gamma\rangle}+\frac{u^{2}2s}{2!(s+1)\langle\Gamma\rangle^{2}}-\frac{u^{3}2s(3s-1)}{3!(s+1)^{2}
\langle\Gamma\rangle^{3}}+...\rightarrow1, \quad
\langle\Gamma\rangle\rightarrow\infty,
$$
as in (\ref{sum}). By finite values of ratio
$s=D/\langle\Gamma^{2}\rangle$ limiting behaviour of NSO is same,
as in (\ref{sum}) and \cite{ry07}.

4c). Let's consider one more distribution for $p_{q}(u)$,
connected with degree laws. In work \cite{fe} by consideration of
the renewal theory it is received distribution for length of an
interval $t-S_{N_{t}}$, where $S_{N}$ are the renewal moments;
$S_{N_{t}}<t<S_{N_{t}+1}$. If to interpret the renewal moments as
a birth and destruction of system the interval  $t-S_{N_{t}}$
represents time of past life of system, the value $t-t_{0}$ from
(\ref{NSO}). In \cite{fe} it is shown, that $P\{t-S_{N_{t}}>x,
S_{N_{t}+1}-t>y\}\rightarrow\mu^{-1}\int_{x+y}^{\infty}(1-F(s))ds$,
where $F(s)$ is distribution of random variable $T_{i}$ from sum
$S_{n}=S_{0}+T_{1}+ ... +T_{n},
\mu=\int_{0}^{\infty}(1-F(s))ds=\int_{0}^{\infty}sF(s)ds$. At
large $s$: $1-F(s)\sim s^{-\alpha}L(s), 0<\alpha <2, L(s)$ is
slowly varying function. Thus, distribution $p_{q}(u)$ at large
values $t$ and $x+y$ looks like $u^{-\alpha}$. We will break all
time interval on two parts and we will describe at large times
function $p_{q}(u)$ degree dependence, and on small times we set
$p_{q}(u)$ gamma function of the form (\ref{gam}), (\ref{ga}) with
$k=2$, i.e (\ref{exp2}). Thus,

\begin{equation}
p_{q}(u)= \{\left.
\begin{array}{l}
\varepsilon^{2}u\exp\{-\varepsilon u\},  \quad u <c;
\\
bu^{-\alpha}, \quad u\geq c,
\end{array}
\right. \label{c}
\end{equation}
where $c$ is some value of
time. From a normalization condition of distribution (\ref{c}) we
find, that at $1 <\alpha <2$,

$$
b=\frac{-(1-\alpha)\varepsilon\exp\{-\varepsilon
c\}(c+1/\varepsilon)}{c^{-\alpha+1}},
$$
But value $\langle\Gamma\rangle$ disperses. Therefore we will be
limited not to an infinite limit of integration on time, and some
limiting value of lifetime $\Gamma_{m}$. In this case the
normalization condition gives value

$$
b=\frac{-(1-\alpha)\varepsilon\exp\{-\varepsilon
c\}(c+1/\varepsilon)}{(\Gamma_{m}^{-\alpha+1}-c^{-\alpha+1})},
$$
and for average value of lifetime we receive expression

\begin{equation}
\langle\Gamma\rangle=\frac{2}{\varepsilon}+\exp\{-\varepsilon
c\}[\frac{(1-
\alpha)\varepsilon(c+1/\varepsilon)(\Gamma_{m}^{-\alpha+2}-c^{-\alpha+2})}{(2-
\alpha)(\Gamma_{m}^{-\alpha+1}-c^{-\alpha+1})}-\frac{(c^{2}\varepsilon^{2}+2c\varepsilon+2)}{\varepsilon}].
\label{c1}
\end{equation}

If to fix parameters $\Gamma_{m}, c, \varepsilon$, dependence
$\varepsilon(\langle\Gamma\rangle)$, defined from (\ref{c1}), will
be positive for a limiting case interesting us

$$
\Gamma_{m}\rightarrow\infty, \quad
\langle\Gamma\rangle\rightarrow\infty, \quad  r_{m}=lim
\langle\Gamma\rangle/\Gamma_{m}
$$
at enough small values $r_{m}$. For example, in case of extension
$\exp\{-\varepsilon c\}$ in series and restrictions of cubic
items, we receive, that

$$
\varepsilon^{2}\approx\frac{(\langle\Gamma\rangle-g)}{c^{2}
(c/3-g/2)};\quad
g=\frac{(1-\alpha)(\Gamma_{m}^{-\alpha+2}-c^{-\alpha+2)}}{(2-\alpha)
(\Gamma_{m}^{-\alpha+1}-c^{-\alpha+1})}.
$$
In a limiting case
$$\Gamma_{m}\rightarrow\infty,\quad
\langle\Gamma\rangle\rightarrow\infty,\quad
\varepsilon^{2}=\frac{2(g_{1}-r_{m})}{c^{2}g_{1}}, \quad
g_{1}=\frac{(1-\alpha)}{(\alpha-2)}>0.
$$
That was $\varepsilon^{2}> 0$, should be $r_{m}$ less $g_{1}$. The
ratio $r_{m}$ is finite at $\Gamma_{m}\sim u^{-\alpha+2}$ as the
value $\langle\Gamma\rangle$ disperses as $u^{-\alpha+2}$ at
$u\rightarrow\infty$. Thus in a limiting case the value
$\varepsilon$ remains finite, and all additives entering in
$p_{q}(u)$ and in additional expressions to NSO all additives are
finite too unlike (\ref{sum}).

If to fix parameters $\Gamma_{m}, c, \varepsilon$, defining
dependence $\alpha$ from $\langle\Gamma\rangle$ from (\ref{c1}) it
is received, that at finite values $c$ and $\varepsilon$, at
$\Gamma_{m}\rightarrow\infty,
\langle\Gamma\rangle\rightarrow\infty, r_{m}=lim
\langle\Gamma\rangle/\Gamma_{m}$

$$
\alpha=\frac{(m+2r_{m})}{(m+r_{m})}, \quad  m=\exp\{-\varepsilon
c\}(\varepsilon c+1).
$$
It is possible to consider and other limiting cases, and other
distributions for $p_{q}(u)$. But the general conclusion consists
in that, as for infinitely large systems and infinitely large
lifetimes the task of realistic distributions for time the lived
of system a life changes a non-equilibrium state of system. The
account of character of history of system, features of its
behaviour in the past, can make essential influence on
non-equilibrium behaviour of system in present time situation.

4d). One more distribution for $p_{q}(u)$ can be received from
results of works [31], integrating distribution $P(E, \Gamma)$ on
$E$

\begin{equation}
p_{q}(u)=\exp\{-\gamma u\}(1-c\exp\{-\gamma u\})^{-1/(q-1)}; \quad
c=(q-1)\gamma a q^{-1};
\label{ryaz}
\end{equation}

$$
a=(\exp\{\beta PV\}-1)^{-1}; \quad  \beta=1/kT,
$$
where $V=\Delta$ is volume of metastable area, $P$ is pressure,
$T$ is temperature \cite{ryaz}. The normalization of distribution
(\ref{ryaz}) and its moments are expressed through incomplete
beta-function \cite{abr}. For example, average value of lifetime
is equal

$$
\gamma
a\langle\Gamma\rangle=\frac{a^{-1}\Gamma^{2}(1/a)_{3}F_{2}(a^{-1},a^{-1},1/(q-1);1+a^{-1},1+a^{-1},c)}{_{2}F_{1}
(a^{-1},1/(q-1);1+a^{-1},c)}=\\
$$
$$
=\frac{a^{-1}\Gamma^{2}(a^{-1})(1+\frac{a\gamma}{q(1+a)^{2}}+\frac{a^{2}\gamma^{2}}{2q(1+2a)^{2}}+...)}{(1+
\frac{a\gamma}{q(1+a)}+\frac{a^{2}\gamma^{2}}{2q(1+2a)}+...)},
$$
$\Gamma(a^{-1})$ is gamma function, $_{n}F_{m}$ is
hypergeometrical function [32]. The ratio of a dispersion to a
square of average value is equal

$$
D/\langle\Gamma\rangle^{2}=
$$
$$
=\frac{2(a)_{4}F_{3}(\frac{1}{a},\frac{1}{a},\frac{1}{a},\frac{1}{(q-1)};1+\frac{1}{a},1+\frac{1}{a},1+\frac{1}{a};c)
_{2}F_{1}(\frac{1}{a},1/(q-1);1+\frac{1}{a};c)}{\Gamma(\frac{1}{a})_{3}F_{2}^{2}(\frac{1}{a},\frac{1}{a},\frac{1}{(q-1)};
1+\frac{1}{a},1+\frac{1}{a};c)}.
$$

If to assume a little values $\gamma$ and $\gamma a$, and to be
limited linear items,

$$
\frac{\gamma}{a}=(\frac{s\Gamma(a^{-1})}{2a}-1)\frac{(1+a)^{3}}{a^{3}};
\quad D/\langle\Gamma\rangle^{2}=s\approx\frac{2a(1+\frac{\gamma
a^{3}}{q(1+a)^{3}})}{\Gamma(1/a)};\\
$$
$$
\frac{\gamma
a^{2}\langle\Gamma\rangle}{\Gamma^{2}(1/a)}\approx1-\frac{a^{2}\gamma}{(1+a)^{2}q}=
1-\frac{(1+a)(\frac{s\Gamma(1/a)}{2a}-1)}{a}.
$$

From here $\gamma\rightarrow0$ at
$\langle\Gamma\rangle\rightarrow\infty$ at finite values $a$, and
we receive Zubarev's result (\ref{NSO}), (\ref{expDi}),
(\ref{sum}) and \cite{ry07}, when additives to NSO becomes
vanishingly small with $\langle\Gamma\rangle\rightarrow\infty$.
The same gives also square-law approach on $\gamma a$.

Let's consider now lifetime distributions of a various form to
various time scales as (\ref{c}). In work \cite{ry07} it was
marked that during evolution the system passes various stages
(kinetic, hydrodynamic, etc.). Lifetime can end at any stage. At
different stages the functions  $\varrho_{q}$ accept a various
kind. Therefore and expression for NSO (\ref{NSO}) becomes
complicated.

4e). Let's consider distribution of kind

\begin{equation}
p_{q}(u)= \big\{\left.
\begin{array}{l}
\varepsilon\exp\{-\varepsilon u\}, \quad u<c;
\\
\frac{bka^{k}}{(u+a)^{k+1}}, \quad u\geq c,
\end{array}
\right.
\label{e}
\end{equation}
combining exponential distribution (\ref{expDi}) for small times
and fractional Pareto distribution (\ref{Pa}) for $u\geq c$. From
a normalization condition
$1=\varepsilon\int_{0}^{c}exp\{-\varepsilon
u\}du+b\int_{c}^{\infty}\frac{kdu}{a(1+u/a)^{k+1}}$ it is found a
normalizing constant, $b=c^{k}exp\{-\varepsilon c\}$, and then the
first and second moments of lifetime:

$$
\langle\Gamma\rangle=\frac{1}{\varepsilon}[1-e^{-\varepsilon
c}(1+\varepsilon c)]-e^{-\varepsilon c}a[1+\frac{kc}{(1-k)}];
$$
$$
\langle\Gamma^{2}\rangle=\frac{2}{\varepsilon^{2}}[1-e^{-\varepsilon
c}(1+\varepsilon
c+\frac{\varepsilon^{2}c^{2}}{2})]-e^{-\varepsilon
c}a^{2}[\frac{kc^{2}}{(2-k)}-\frac{2kc}{(1-k)}-1].
$$

For Pareto distribution at $k<2$ the second moment does not exist
\cite{kor}. The value $\langle\Gamma\rangle\rightarrow\infty$ with
$k\rightarrow1$ from above. Limiting transition $k\rightarrow1$
should be spent after thermodynamic limiting transition. The
values $\varepsilon, c, a$ remains finite. Then

$$
-\int p_{q}(u)du= \big \{\left.
\begin{array}{l}
\exp\{-\varepsilon u\}, \quad u<c;
\\
\frac{c^{k}\exp\{-\varepsilon c\}}{(1+u/a)^{k}}, \quad u\geq c
\end{array}
\right.;
$$
$$
-\int p_{q}(u)du
\left.
\begin{array}{l}
\Rightarrow
\\
k\rightarrow 1
\end{array}
\right. \big\{\left.
\begin{array}{l}
(1-\varepsilon u+...), \quad u<c;
\\
c\exp\{-\varepsilon c\}(1-u/a+...), \quad u\geq c.
\end{array}
\right.
$$

Thus, in this case additives to unit and to Zubarev's NSO are not
equal to zero and for infinitely large systems with
$\langle\Gamma\rangle\rightarrow\infty$. Uncertain there are
values of parameters  $\varepsilon, c, a$. As function $p_{q}(u)$
it is possible to choose both more simple and more difficult
functions.

4f). If to choose

$$
p_{q}(u)= \big\{\left.
\begin{array}{l}
\varepsilon\exp\{-\varepsilon u\}, \quad u<c;
\\
b, \quad u\geq c,
\end{array}
\right.
$$
that $b=\frac{\exp\{-\varepsilon c\}}{(\Gamma_{m}-c)}$, integral
in limits from $0$ to $\Gamma_{m}$ , but not from $0$ to $\infty$.

$$
\langle\Gamma\rangle=\frac{1}{\varepsilon}[1-e^{-\varepsilon
c}(1+\varepsilon c)]+e^{-\varepsilon c}\frac{(\Gamma_{m}+c)}{2};
$$
$$
\langle\Gamma\rangle\rightarrow\infty, \quad
\Gamma_{m}\rightarrow\infty, \quad
\frac{\langle\Gamma\rangle}{\Gamma_{m}}\rightarrow\frac{e^{-\varepsilon
c}}{2};
$$
$$
\langle\Gamma^{2}\rangle=\frac{2}{\varepsilon^{2}}[1-e^{-\varepsilon
c}(1+\varepsilon
c+\frac{\varepsilon^{2}c^{2}}{2})]+\frac{e^{-\varepsilon
c}}{3}(\Gamma_{m}^{2}+\Gamma_{m}c+c^{2});
$$
$$
\frac{\langle\Gamma^{2}\rangle}{\Gamma_{m}^{2}}\rightarrow\frac{e^{-\varepsilon
c}}{3}=\frac{2\langle\Gamma\rangle}{3\Gamma_{m}}, \quad
\Gamma_{m}\rightarrow\infty.
$$

If after thermodynamic limiting transition
$\Gamma_{m}\rightarrow\infty$, then $b\rightarrow0$,

$$
p_{q}(u)= \big\{\left.
\begin{array}{l}
\varepsilon\exp\{-\varepsilon u\}, \quad u<c;
\\
\frac{\exp\{-\varepsilon c\}}{(\Gamma_{m}-c)}, \quad u\geq c;
\end{array}
\right.
$$
$$
p_{q}(u)\Rightarrow  \big\{\left.
\begin{array}{l}
\exp\{-\varepsilon u\}, \quad u<c; \\
0, \qquad u\geq c.
\end{array}
\right.
$$

The value $\varepsilon$  is finite, and additives to unit are not
equal to zero. But in this case the effect of "finite memory",
limited on time by the size $c$, is observed.

4g). For

\begin{equation}
p_{q}(u)= \big\{\left.
\begin{array}{l}
\varepsilon_{1}^{2}u\exp\{-\varepsilon_{1} u\}, \quad u<c;
\\
b\varepsilon_{2}\exp\{-\varepsilon_{2} u\}, \quad u\geq c,
\end{array}
\right.
\label{g}
\end{equation}
we write down from a normalization condition
$b=\exp\{-\varepsilon_{1}c\}(1+\varepsilon_{1}c)$ and the two
first moment

$$
\langle\Gamma\rangle=\frac{1}{\varepsilon_{2}}e^{-\varepsilon_{1}c}(1+\varepsilon_{1}c)e^{-\varepsilon_{2}c}
(1+\varepsilon_{2}c)+\frac{2}{\varepsilon_{1}}[1-e^{-\varepsilon_{1}c}(1+\varepsilon_{1}c+\frac{(\varepsilon_{1}c)^{2}}{2})];
$$

$\langle\Gamma\rangle\rightarrow\infty$ by
$\varepsilon_{2}\rightarrow0$;

$$
\langle\Gamma^{2}\rangle=\frac{6}{\varepsilon_{1}^{2}}[1-e^{-\varepsilon_{1}
c}(1+\varepsilon_{1}c
+\frac{(\varepsilon_{1}c)^{2}}{2}+\frac{(\varepsilon_{1}c)^{3}}{6}]+
$$
$$
+e^{-\varepsilon_{1}c}\frac{2}{\varepsilon_{2}^{2}}
e^{-\varepsilon_{2}c}(1+\varepsilon_{1}c)(1+\varepsilon_{2}c+\frac{(\varepsilon_{2}c)^{2}}{2}).
$$

Then

$$
-\int p_{q}(u)du= \big\{\left.
\begin{array}{l}
(1+\varepsilon_{1}u)e^{-\varepsilon_{1}u}, \quad u<c;
\\
(1+\varepsilon_{1}c)e^{-\varepsilon_{1}c}e^{-\varepsilon_{2}u},\quad
 u \geq c;
\end{array}
\right.
$$
$$
-\int p_{q}(u)du\left.
\begin{array}{l}
\Rightarrow
\\
\varepsilon_{2}\rightarrow0, \langle\Gamma\rangle\rightarrow\infty
\end{array}
\right.
\big\{\left.
\begin{array}{l}
1-\frac{(\varepsilon_{1}u)^{2}}{2}+..., \quad u<c;
\\
(1+\varepsilon_{1}c)e^{-\varepsilon_{1}c}, \quad u \geq c.
\end{array}
\right.
$$

Additives to unit to tend to infinity of average lifetime are not
equal to zero.

4h). For

$$
p_{q}(u)= \big\{\left.
\begin{array}{l}
\varepsilon^{2}u\exp\{-\varepsilon u\}, \quad u<c;
\\
\frac{b\exp\{-\gamma u\}}{[1+(q-1)a\gamma\exp\{-\gamma
au\}/q]^{1/(q-1)}}, \quad u\geq c,
\end{array}
\right.
$$
combination of (\ref{exp2}) and (\ref{ryaz}),

$$
b=\frac{\gamma a[(q-1)\gamma a/q]^{1/a}[1-\exp\{-\varepsilon
c\}(1+\varepsilon c)]}{B_{(1-p,1)}(1-1/(q-1),1/a)}; \quad
p=(q-1)a\gamma\exp\{-\gamma ac\}/q ;
$$
$$
\langle\Gamma\rangle=\frac{2}{\varepsilon}[1-\exp\{-\varepsilon
c\}(1+\varepsilon c+\frac{(\varepsilon c)^{2}}{2})]+
$$
$$
+[1-\exp\{-\varepsilon c\}(1+ \varepsilon
c)]\frac{\Gamma^{2}(1/a)_{3}F_{2}(1/a,1/a,1/(q-1);1+1/a,1+1/a;p)}{(a^{2}\gamma)
_{2}F_{1}(1/a,1/(q-1);1+1/a;p)};
$$
$B_{(1-p,1)}(1-1/(q-1),1/a)$ is incomplete beta function;
$\langle\Gamma\rangle\rightarrow\infty$ when $\gamma\rightarrow
0$, as in 4d), (\ref{ryaz}), $b\rightarrow 0$ when
$\gamma\rightarrow 0$, after of thermodynamic limiting transition,
and

$$
p_{q}(u)du\left.
\begin{array}{l}
\Rightarrow
\\
\gamma\rightarrow 0
\end{array}
\right. \big\{\left.
\begin{array}{l}
\varepsilon^{2}u\exp\{-\varepsilon u\}, \quad u<c;
\\
0, \quad u\geq c.
\end{array}
\right.
$$
$$
-\int p_{q}(u)du\left.
\begin{array}{l}
\Rightarrow
\\
\gamma\rightarrow 0
\end{array}
\right. \big\{\left.
\begin{array}{l}
\varepsilon\exp\{-\varepsilon u\}(u+1/\varepsilon), \quad u<c;
\\
0, \quad u\geq c.
\end{array}
\right.
$$
Thus, as in 4f) it is received "finite memory" on an interval $(0,
c)$, but contributions in NSO, amendments to Zubarev's NSO are
finite and at $\langle\Gamma\rangle\rightarrow\infty$. "Finite
memory" is possible and for "usual" functions of distribution by
$u\geq c$ and for fractional distributions of type of Pareto
distribution. So, for

4i).

$$
p_{q}(u)du= \big \{\left.
\begin{array}{l}
\varepsilon\exp\{-\varepsilon u\}, \quad u<c;
\\
\frac{b\exp\{-\gamma u\}}{(1+au)^{m}}, \quad u\geq c,
\end{array}
\right.
$$
from a normalization condition

$$
b=\frac{a\exp\{-\varepsilon c\}\exp\{-\gamma
a\}}{(\gamma/a)^{1-m}\Gamma(1-m, \gamma/a)},
$$
$\Gamma( , )$ is incomplete gamma function,

$$
\langle\Gamma\rangle=\frac{1}{\varepsilon}[1-\exp\{-\varepsilon
c\}(1+\varepsilon c)]+\exp\{-\varepsilon
c\}[\frac{\gamma\Gamma(2-m, \gamma/a)}{a^{2}\Gamma(1-m,
\gamma/a)}-1].
$$
The first moment $\langle\Gamma\rangle\rightarrow\infty$ at
$a\rightarrow 0$. In this case $b\rightarrow\gamma$, and

$$
p_{q}(u)du\left.
\begin{array}{l}
\Rightarrow
\\
a\rightarrow 0
\end{array}
\right. \big\{\left.
\begin{array}{l}
\varepsilon \exp\{-\varepsilon u\}, \quad u<c;
\\
\gamma\exp\{-\varepsilon c\}\exp\{-\gamma u\}, \quad u\geq c.
\end{array}
\right.
$$
$$
-\int p_{q}(u)du= \big \{\left.
\begin{array}{l}
\exp\{-\varepsilon u\}, \quad u<c;
\\
-\gamma\exp\{-\varepsilon c\}\int\frac{\exp\{-\gamma
u\}du}{(1+au)^{m}}, \quad u \geq c;
\end{array}
\right.
$$
$$
-\int p_{q}(u)du\left.
\begin{array}{l}
\Rightarrow
\\
a\rightarrow 0
\end{array}
\right. \big\{\left.
\begin{array}{l}
\exp\{-\varepsilon u\}, \quad u<c;
\\
\exp\{-\gamma u\}\exp\{-\varepsilon c\}, \quad u\geq c,
\end{array}
\right.
$$
as in a case (\ref{e}), i.e. additives do not address in a zero at
$u\geq c$. The value $\langle\Gamma\rangle\rightarrow\infty$ and
at $\gamma\rightarrow\infty$. In this case $p_{q}(u)=0$ at $u>c$,
memory is finite.

\section{The conclusion}
 \label{sect:4}

As it is specified in work \cite{rau}, existence of time scales
and a stream of the information from slow degrees of freedom to
fast create irreversibility of the macroscopical description. The
information continuously passes from slow to fast degrees of
freedom. This stream of the information leads to irreversibility.
The information thus is not lost, and passes in the form
inaccessible to research on Markovian level of the description.
For example, for the rarefied gas the information is transferred
from one-partial observables to multipartial correlations. In work
\cite{ry01} values  $\varepsilon=1/\langle\Gamma\rangle$ and
$p_{q}(u)=\varepsilon\exp\{-\varepsilon u\}$ are expressed through
the operator of entropy production and, according to results
\cite{rau}, - through a stream of the information from relevant to
irrelevant degrees of freedom. Introduction in NSO function
$p_{q}(u)$ corresponds to specification of the description by
means of the effective account of communication with irrelevant
degrees of freedom. In the present work it is shown, how it is
possible to spend specification the description of effects of
memory within the limits of method NSO, more detailed account of
influence on evolution of system of quickly varying variables
through the specified and expanded kind of density of function of
distribution of time the lived system of a life.

In many physical problems finiteness of lifetime can be neglected.
Then $\varepsilon\sim1/\langle\Gamma\rangle\rightarrow0$. For
example, for a case of evaporation of drops of a liquid it is
possible to show \cite{ryDr}, that non-equilibrium characteristics
depend from $\exp\{y^{2}\}; y=\varepsilon/(2\lambda_{2})^{1/2},
\lambda_{2}$ is the second moment of correlation function of the
fluxes averaged on quasi-equilibrium distribution. Estimations
show, what even at the minimum values of lifetime of drops
(generally - finite size) and the maximum values size
$y=\varepsilon/(2\lambda_{2})^{1/2}\leq 10^{-5}$. Therefore
finiteness of values $\langle\Gamma\rangle$ and $\varepsilon$ does
not influence on behaviour of system and it is possible to
consider $\varepsilon=0$. However in some situations it is
necessary to consider finiteness of lifetime
$\langle\Gamma\rangle$ and values  $\varepsilon> 0$. For example,
for nanodrops already it is necessary to consider effect of
finiteness of their lifetime. For lifetime of neutrons in a
nuclear reactor in work \cite{ry01} the equation for
$\varepsilon=1/\langle\Gamma\rangle$ which decision leads to
expression for average lifetime of neutrons which coincides with
the so-called period of a reactor is received. In work \cite{ryAt}
account of finiteness of lifetime of neutrons result to correct
distribution of neutrons energy.

Use of distributions (\ref{gam}), (\ref{exp2}), (\ref{Pa}),
(\ref{ryaz}) and several more obvious forms of lifetime
distribution in quality $p_{q}(u)$ leads to a conclusion, that the
deviation received by means of these distributions values
$\ln\varrho(t)$ from $\ln\varrho_{zub}(t)$ is no more
$\varepsilon\sim1/\langle\Gamma\rangle$. Therefore in expression
(\ref{NSO}) additives to the result received by Zubarev, are
proportional $\varepsilon$. This result corresponds to
mathematical results of the theory asymptotical phase integration
of complex systems \cite{tur} according to which distribution of
lifetime looks like $p_{q}(u)=\exp\{-\varepsilon
u\}+\lambda\varphi_{1}(u)+\lambda^{2}\varphi_{2}(u)+...$ , where
the parameter of smallness $\lambda$ in our case corresponds to
value $\varepsilon\sim1/\langle\Gamma\rangle$. Generally the
parameter $\lambda$ can be any.

For distributions of kind (\ref{c}), (\ref{e}), (\ref{g}), having
a various form for different times, additives to Zubarev's NSO are
distinct from zero and for infinitely large systems with
infinitely large lifetimes. For some distributions the effect of
"finite memory" when only the limited interval of the past
influences on behaviour of system is observed.

Probably, similar results will appear useful, for example, in
researches of small systems. All greater value is acquired by
importance of description of the systems in mesoscopical scales. A
number of the results following from interpretation of NSO and
$p_{q}(u)$ as density of lifetime distribution of system
\cite{ry01}, it is possible to receive from the stochastic theory
of storage \cite{prab} and theories of queues. For example, in
\cite{prab} the general result that the random variable of the
period of employment (lifetime) has absolutely continuous
distribution $p_{q}(u)=g(u,x)=xk(u-x,u), u>x>0$ is resulted;
$g(u,x)=0$ in other cases, where $k(x,t)$ is absolutely continuous
distribution for value $X(t)$ - input to system.

The form of distribution chosen by Zubarev for lifetime represents
limiting distribution. The choice of lifetime distribution in NSO
is connected with the account of influence of the past of system,
its physical features, for the present moment, for example, with
the account only age of system, as in Zubarev's NSO
\cite{zub74,zub80,ry01} at $\varepsilon>0$, or with more detailed
characteristic of the past evolution of system. The received
results are essential in cases when it is impossible to neglect
effects of memory when memory time there is not little. The
analysis of time scales as it is noted in \cite{rau}, it is
necessary to spend in each problem.

Generalization of the received results on wider (generally any
distributions $F(x)$) classes of distributions is received in work
\cite{fe} with use of methods of the renewal theory. In \cite{fe}
it is shown, that the normalizing random variable $(t-t_{0})/t$ at
$t\rightarrow\infty$ has limiting density
$g_{\alpha}(x)=(sin\pi\alpha/\pi)x^{-\alpha}(1-x)^{\alpha-1},
0<\alpha<1, x\in [0,1]$, connected with functions of distribution
$F(x)$, having correctly varying tails, $1-F(x)=x^{-\alpha}L(x),
0<\alpha<1$, where $L(tx)/L(t)\rightarrow1$ at
$t\rightarrow\infty$. Average value
$\langle\Gamma\rangle/t=(\alpha-1)sin\pi\alpha/sin\pi(\alpha-1)$.
As $\langle\Gamma\rangle/t=\delta$ is small size values $\alpha$
are close to unit and $\delta\approx\sin\pi\alpha/\pi$. At
$\alpha\approx1-\delta,
sin(1-\delta)\pi=sin\pi\delta=\delta\pi-(\delta\pi)^{3}/3+...\approx\pi\delta$,
we receive identity. At $\alpha\approx1-\delta$ distribution
$g_{\alpha}(x)\approx\delta(1-x)^{-\delta}/x^{1-\delta}$ behaves
in the similar image with $\varepsilon\exp\{-\varepsilon x\}$ at
$\varepsilon\sim\delta$, differing at $x\rightarrow0$. In this
case universal distribution also is characterized only by one
parameter $\alpha$, but the limiting situation
$t\rightarrow\infty$ and influence of tails of distribution,
probably, not absolutely full describes past influence on the
present as the near moments of time are thus more significant,
with the memory which has not gone yet.

In Prigogine's work \cite{Prig} of function of distribution,
evolving in course of time in accordance with the laws of
mechanics, through transformation a distribution function is put
in accordance, the evolution of which is described by
probabilistic rule. The role of such transformation in the method
of NSO plays averaging on the density of distribution of time by
the spent system of life.

If type of source in Liouville equation for a non-equilibrium
statistical operator in the form of Zubarev \cite{zub80} it is
possible to confront with a linear relaxation source in Boltzmann
equation, more difficult types of sources, got from other
distributions for lifetime of the system, it is possible to
compare to more realistic type of integral of collisions, that is
explained by the openness of the system, by its co-operation with
surroundings and finiteness of lifetime of the system, and also
coarsening for physically infinitely small volumes.

The problems of correlation of spatial and temporal descriptions
are interesting. So, spatial description in works of Klimontovich
\cite{klim}, smoothing out on physically infinitely small volume,
yields to results similar with the results of Zubarev, to the same
expression for a source. Account of influence of surroundings,
interaction with other systems (spatial task) also brings to the
similar results. So, in works of MacLennan, short exposition of
which and accordance with Zubarev's works is given in the Appendix
2 to the book of Zubarev \cite{zub80}, a source in Liouville
equation become formed by thermodynamical variables - temperature,
chemical potential and speed characterizing surroundings, not
details of his microscopic state. If, as in works \cite{ra95,ra99}
to conduct replacement of temporal argument of thermodynamical
variables and to increase them on the proper "weigth functions"
(which in \cite{zub74} is interpreted as densities of distribution
of time by the spent system of life), we will get stated in the
offered work results.

Correlations between the exponential damping (\ref{expDi}) and
nonexponential functions just for small, non-Markov time scale
factors, considered for a quantum case in works \cite{prig01}.

\end{document}